\documentclass[conference]{IEEEtran}

\usepackage{eso-pic}
\newcommand\AtPageUpperMyright[1]{\AtPageUpperLeft{%
 \put(\LenToUnit{0.5\paperwidth},\LenToUnit{-1cm}){%
     \parbox{0.5\textwidth}{\raggedleft\fontsize{9}{11}\selectfont #1}}%
 }}%
\newcommand{\conf}[1]{%
\AddToShipoutPictureBG*{%
\AtPageUpperMyright{#1}
}
}

\ifCLASSINFOpdf
  \usepackage[pdftex]{graphicx}
  \DeclareGraphicsExtensions{.pdf,.jpeg,.png}
\else
\fi

\usepackage[cmex10]{amsmath}

\usepackage{url}

\begin{document}
\title {Decaying Indicators of Compromise}

\author{\IEEEauthorblockN{Andras Iklody\\ G\'erard Wagener  \\ Alexandre Dulaunoy  and \\ Sami Mokaddem}
\IEEEauthorblockA{\\
CIRCL- Computer Incident Response Center Luxembourg \\
Email: info@circl.lu\\
}
\and
\IEEEauthorblockN{\\ Cynthia Wagner}\\
\\
\IEEEauthorblockA{Fondation RESTENA\\
CSIRT\\
Email: cynthia.wagner@restena.lu}}

\maketitle
\conf{DRAFT 1.0 - for public review - feedback info@circl.lu}

\begin{abstract}

The steady increase in the volume of indicators of compromise (IoC) as well
as their volatile nature makes their processing challenging.
Once compromised infrastructures are cleaned up, threat actors are moving to
on to other target infrastructures or simply changing attack strategies.
To ease the evaluation of IoCs as well as to harness the combined analysis
capabilities, threat intelligence sharing platforms were
introduced in order to foster collaboration on a community level.
In this paper, the open-source threat intelligence platform MISP is used to
implement and showcase a generic scoring model for decaying IoCs shared
within MISP communities matching their heterogeneous objectives. The model takes into
account existing meta-information shared along with indicators of compromise,
facilitating the decision making process for machines in regards to the validity of
the shared indicator of compromise. The model is applied on common use-cases that are
normally encountered during incident response.

\end{abstract}


%
\IEEEpeerreviewmaketitle


\section{Introduction}
Information sharing platforms nowadays have become an important tool in fighting
active threats and a precious source of information about the actual
indicator of compromise lanscape. One of these platforms is
 MISP \cite{misp-article}, an open source threat intelligence sharing
platform that enables various actors from private or public IT-communities
to share a wide range of their information, may this be IoCs, malware or other
relevant data about existing threats.
A particularity of MISP is that it broke with the traditional 'single producer
multiple consumer' paradigm that can commonly be observed in other threat
intelligence platforms and rather implements a peer to peer architecture. Due to
this, each participating member can produce, enhance or consume information
and give feedback on information produced by others. Pieces of information can
be transfered along multiple nodes between partners. Hence, the data
quality and the trust of the data source is not always certain.
To provide a reliability and credibility measure to data providers,
some data interaction models are applied, as for example taxonomies.

By giving data context, such as descriptions or adding attributes, raw data
becomes precious information for others. Data is volatile and thus becomes
outdated or even invalid, as for example, compromised machines get cleaned up or
reused, IP addresses can change or domain names can be deleted. Therefore a model
is proposed which uses the existing shared pool of data, matching the various, diverse
objectives of community members, such as operational network security or
threat intelligence along with community defined metrics and classifications to
scope the lifecycle of IoCs for various use-cases.

The paper is organized as follows, section \ref{related} discusses relevant approaches focussing on threat intelligence collection, processing and sharing. Section \ref{misp-tool} introduces to MISP and describes this platform in general and its data interaction models. In section \ref{sighting} attribute scoring methods are introduced to the reader. Since the research is still ongoing, some future work and conclusions will be presented in section \ref{conclusions}.


\section{Related work}
\label{related}

Sharing indicators of compromise within a community can have a direct impact
on the reaction times to an actual threat. Research in cybersecurity
shows that information sharing within a community is one of the key factors to
accurate incident response, as shown in the study of \cite{Enisa}.
Recurring problems with information sharing include the fact that it is a collective
effort based on a give-and-take approach \cite{Haass2015}. Another major concern is 
the quality of information, especially when it comes to achieving low false positive
rates. Having reliable information is a major concern in information sharing as
described in article \cite{Dandurand13}, where a concept of knowledge management is introduced
to assess requirements for information sharing tools. The authors of
\cite{Pinto2015} describe similar requirements by identifying challenges
for threat intelligence platforms, as for example privacy or quality control
approaches. For this, the authors of article \cite{Maasberg2016} introduce
an assessment approach for malware threat levels referring to scoring and
weighting factors. Other authors, such as those in article \cite{Woods2015}
apply a data mining approach to identify similarity metrics in statistical
relations for shared information. A data-driven visualisation approach is
presented in article \cite{Adams2011}, which evaluates content from news and
social media based on emotions to increase the added value of
information.

Event detection and evaluation are complex tasks, where in intrusion
detection it is often referred to as threshold-based methods for triggering
alarms, such as presented in articles \cite{Prasad} \cite{Mitchell}.
Threshold-based detection are considered as reliable and is often applied
in statistics, data mining or game-theoretical evaluation approaches
\cite{Barford} \cite{Ghafouri} \cite{blinc} \cite{Nikulin}. Beside the
evaluation techniques of data, the data itself provides a large amount of
information, such as IP-addresses, protocols, timestamps, etc,
that play a major role in information sharing and a lot of evaluation
techniques are investigated \cite{garcia} \cite{Mun} \cite{Waldvogel}.

In the area of IoC tracing and evaluation, various approaches exist,
from analysing technical articles or blog entries to the deep analysis of
samples to extract IoC information. Technical focus is given in \cite{Dridex},
where malware samples are evaluated in order to output IoCs by analysing
traffic information.  In article \cite{igen} IoCs are automatically extracted from
different sources, such as reports, articles, etc. and evaluated using
convolutional neural networks to correlate data with other indicators in
order to set up rules to be deployed in a network.
A similar idea was presented in article \cite{Acing} that studies technical articles and
blog entries, but here, natural language processing schemes are applied
to evaluate data and to identify decay times for observed IoCs in text
appearances.

\section{Background and data interaction techniques  in MISP}
\label{misp-tool}

The MISP software is introduced in this section with the focus on the
features used in the scoring model presented in section \ref{sighting}.
For a detailed description of the design and implementation of MISP,
we refer the reader to article \cite{misp-article}. MISP is an open-source threat
information sharing platform, where information on all kinds of threats can be shared
within communities or subsets of community members. Such pieces of information can typically
include indicators of compromise such as IP addresses and file hashes, but also other types of indicators
such as financial indicators can be shared within communities or a subset of them.

Users can decide on the granularity of
information they wish to disclose in MISP by sharing only some subsets of the
information package. The sharing level can also be set, using the following levels:
organisation only, the current community, directly inter-connected communities,
managed distribution lists via "sharing groups", or simply
all available communities. With the distribution level set to organisation only,
the information is kept exclusively for the organisation of the information producer.
Community only allows the sharing of information among all members of the given MISP instance.
Data marked as connected communities will be made available to all users of the given MISP
instance along with all members of any MISP that has a direct link to the current instance.
For more complex sharing scenarios, sharing groups allow users to create reusable or ad hoc groups,
including a list of defined, involved partners.
Popular examples for sharing groups are organisations grouped by
business sectors such as actors within the telecom sector or financial sector.
Information can also travel through $N$ hops such as the connected communities
and their connected communities etc when the sharing level is set to all.
Hence, MISP uses a peer to peer architecture. Three methods of
synchronization exist between connected MISP instances: push, pull and
cherry pick.  In the last method, administrators can manually pick the events from a
connected MISP instance they want to share within their own community.

In MISP shared events can be populated with one of 140 different types of attributes (destination IP addresses, file hashes)
as well as a list of community developed object templates, combining clusters of linked attributes into logical containers.

Attributes themselves can be defined as a tuple of (category, type, value), conveying both
direct actionable data as well as its associated context.
Additional contextual information such as the date, threat level, description,
organisation, and higher level information such as that on threat actors among others can also be attached
to events. 
Consumers have the possibility to create proposals, which pending the producer's validation can become attributes.
Communication between participants can happen through the built in discussion system.
Events can be filtered according to the various taxonomies described via the
standard format defined in the IETF document \cite{Dulaunoy}.

A taxonomy in MISP is based on the machine-tag approach with
triple-tags for representing semantic information. The method was introduced by
Flickr for the geolocation of pictures \cite{straup2007machine}.
The triple-tag syntax is a simple expression that has a namespace, a predicate
and a value, as shown  in the following example: \{\textit{nato} :
\textit{classification} = \textit{'NU'}\}, this means that \textit{nato} is the
namespace, \textit{classification} is the predicate and 'NU' the value.

The public repository\footnote{https://github.com/MISP/misp-taxonomies} with
MISP taxonomies includes 47 different taxonomies for the domains of law
enforcement, computer security incident response team (CSIRT)
classifications, intelligence and many more. Each community can define their
own taxonomy explaining the growing of used taxonomies.
Intrusion detection systems can also import the latest signatures from MISP and integrate them in their detection rule set. Some intrusion detection systems can be instrumented to sent a REST (REpresentational State Transfer) request towards MISP with feedback about its detections. This feedback is called \texttt{sighting}. Thus, knowledge can be gathered about the validity, freshness of an information or its impact. A typical example is an IP address of a compromised website distributing malware or acting as command and control server which is cleaned up after a while. In MISP the concept of detection and sighting is applicable for almost all kind of attributes. Hence, it can be used by host intrusion detection systems capable of detecting malicious files.  Accounting software was also observed capable of
fetching bank accounts of money mules from MISP for warning accountants in case
of wire transfers towards those. These automated mechanisms are possible due
to the API (Application Programming Interface)
that is heavily used by third parties. At the moment of writing 28 known
tools are capable of interacting with MISP such as Splunk, McAfee,
TheHive\footnote{\url{http://www.misp-project.org/tools/}}.

The main reason to present the approach of sightings and taxonomies in this
section is that they play a major role as parameters in the scoring
model presented in section \ref{sighting}.




\section{Scoring Indicators of Compromise}
\label{sighting}
To illustrate the challenges within MISP communities and the need for a scoring model for attributes, this section will highlight some
numbers extracted from our community we operate for the
private sector \cite{misppriv}.

In its early operation in 2012, the users
knew each other and the trust and data quality was granted in an implicit
way. In January 2018, this community contains 1646 users
from 845 organizations and it is almost impossible that each users know each other, such that there is no explicit trust.
The more, the objectives from each user are different. Users consuming events for
operational security including blocking actions do not want to have false
positives. People performing threat intelligence activities by correlating
indicators of various threat actors want to have a maximum of indicators and
even want information about false positives as they give indications about the
dynamics between theat actors and defenders. Hence, these organizations need
reliable historical data. The correlation feature of MISP is a major driving
incentive for sharing information. Producers see whether other organizations
encountered the same threats.
At the time of writing, the users shared 8686 events having 1048405 attributes. 260896 correlations are found giving the users incentives to share information. They can find out if other organizations were also targeted. The data interaction methods start to grow as 54347 proposals are active and 407 discussion posts are created.
This community includes 8  sighting sources ranging from incident response systems to honeypots giving both indications about the freshness of information.
In total 10219 sightings were recorded including 30 confirmed false positives.

The lifetime of the various available attributes are not homogeneous. For
example, machines cleaned, IP addresses changes
up, IP addresses or domain names are traded and get used in different
fashions over time. Hence, each attribute has its own decay function. File
hashes usually tend not to vary over time. Nevertheless, a shared file hash
can be declared as false positive over time by organizations with distinct
trust and knowledge levels. An example a legitimate file that was embedded in
a malware.

\subsection*{The scoring model for IoCs}

Hence, a model of scores per attribute is selected including the
following conditions:
\begin{itemize}
    \item The base score of an attribute ($a$), called $base\_score_{a}$, is
           a weighting of the confidence of its source and  its linked
          taxonomies ($x$). It is the initial value of the life cycle
          of an indicator. To this value, the score is reset upon a new sighting.
    \item The elapsed time an attribute was seen first and seen last.
    \item The end-time of an attribute $\tau_a$ represents the time at which
          the overall score should be 0.
    \item The \texttt{decay rate} $\delta_a$ represents the speed at which the overall
          score is decreasing over time. The decay speed is variable over time
          as motivated in the following example:\\
          The decay rate of the IP should be low in
          the first hours, but should go faster the more time passes. The first
          time activities from this IP are sighted, the better chances are
          that the threat actors are still active or are executing follow up
          operations. When this IP address is shared among a community targeted
          by the threat actors, more and more members can take measures, such as
          blocking the IP address. Hence, the attack becomes ineffective
          forcing threat actors to use other IP addresses. In case, the IP
          is given up, it could be reassigned to a legitimate customer of the
          Internet service provider leading to collateral damage due to the
          blocking actions of this IP.
\end{itemize}
Information about threats are produced in a collaborative manner
using the data annotation and interaction techniques described in section \ref{misp-tool}.
Tags can be added to events by producers and are defined in a
taxonomy. Some taxonomies enable to express their confidence or
reliability of a  source regarding a  given piece of information they are attached to.
Consumers get this information and have different levels of trust in the producers.

The {$base score_{a}$} for an attribute is defined in equation \ref{basescore},
 $base\_score \in \left[ 0, 100 \right]$. It represents the score of an attribute before taking into account its decay. It is composed of its weighted applied tags and its source confidence.

The weights of the applied taxonomies are defined at
predicate level of each taxonomy and represent its acceptance within a
community. For instance, if tags from the taxonomy with the namespace
{\tt admiralty-scale} and with the predicate {\tt source-reliability} are
hardly used, it gets a low weight. However, if within the same taxonomy tags
with the predicate {\tt information-credibility} are regularly used, it gets
a higher weight.

The \textit{source confidence}
can also be influenced by an additional parameter called $\omega_{sc}$. This
parameter takes into account more subtle trust evaluations.
For example, it could be that an organization has a good image and a
good reputation but due to some circumstances within a given time frame,
the trust in this organization is decreased. A practical example is an
organization that was compromised or taken over by the attacking party.

\begin{equation}\label{basescore}
    base\_score_{a} = weight_{x} \cdot tags + \omega_{sc} \cdot
    source\_confidence
\end{equation}

The $base\_score_{a}$ is defined in equation \ref{basescore} with,

\begin{itemize}
    \item $\forall weight_x \in \left[0, 100\right], \forall \omega_{sc} \in
\left[0, 100\right], weight_{x} + \omega_{sc} = 100$, $weight_x = 100$ or
$\omega_{sc}=100$, a mean to adjust the focus
either on the $tags$ or on the $source\_confidence$. As little research on the trust
rebalancing  and trust evolution of organizations in distributed threat
        sharing is done, the  $\omega_{sc}$ parameter is set to $100-weight_{x}$ and is considered as future work implying further research.
    \item $tags \in \left[0, 1\right]$ is the score derived from the taxonomies
          and is defined in equation \ref{tags}.
    \item $source\_confidence \in \left[0, 1\right]$, is the confidence
          given to the source that published the attribute.
The $source\_confidence$ parameter in equation \ref{basescore} gives a
possibility to influence the $base\_score$, which should be a number
between 0 and 100. Each source between $1$ and $N$ has its $source\_confidence$
level. In case a source is fully trusted the $source\_confidence$ is set to 1.
If there is no trust, the source level is set to 0. The user could also set
intermediate values, which could give an estimate on how reliable the
source is. The learning of the confidence of a source based on its produced
information over time is subject to future research.
\end{itemize}

The relevant taxonomies are summarized in table  \ref{confidence_table}.

The MISP taxonomy includes also a confidence level that is set to `Confidence cannot be evaluated'. This special confidence level cannot be mapped to a numerical value. 
One possibility is to introduce the concept of `undefined'. 

Once a value
is undefined, the $base\_score$ cannot be computed and becomes undefined.
At the end, the overall score would be undefined and by this, cancel
other scoring factors defined in tags. Hence, when the confidence level is
"Confidence cannot be evaluated", it will be ignored.

\begin{center}
\begin{table}
\begin{tabular}{|l|l|l|l|}
    \hline
    Description & Value(s) & Description & Value(s)\\
    \hline
    \multicolumn{2}{|c|}{Misp}&\multicolumn{2}{|c|}{OSINT}\\
    \hline
    Completely confident  & 100 &Certain & 100 \\
    Usually confident & 75& Almost certain & 93\\
    Fairly confident & 50 & Probable&75\\
    Rarely confident & 25 & Chances about even& 50\\
    Unconfident & 0& Probably not&30\\
    &&Impossibility & 0\\
    \hline
\end{tabular}
\vspace{0.3cm}
    \caption{MISP and OSINT taxonomies}
    \label{confidence_table}
\end{table}
\end{center}
\vspace{-0.9cm}
The score derived from the taxonomies is defined in equation \ref{tags}, where
$G$ is the number of defined taxonomy groups and $T$ the number of used
taxonomy per group. The weights are defined at predicate level in the
taxonomies and should be integer numbers between 0 and 100.

\begin{equation}\label{tags}
    tags = \frac{\sum_{j=1}^{j=G} \sum_{i=1}^{i=T} taxonomy_{i} \cdot weight_{i}}{{\sum_{j=1}^{j=G} \sum_{i=1}^{i=T} 100 \cdot weight_{i}}}
\end{equation}

The idea is to decrease the base score over time. When it reaches zero, the
related indicator can be discarded.
A first idea to express the overall score could be
to use  equation \ref{score1}.

\begin{equation}\label{score1}
    \texttt{score}_a = \texttt{base\_score} - \delta_a (T_{t} - T_{t-1})
\end{equation}
where,
\begin{itemize}
    \item $base\_score_{a}$ $\in \left [0, 100\right ]$ is
    described in equation \ref{basescore}.
    \item $\delta_a$ $\in \left [0, +\infty \right.$ represents the decay rate,
          or expressed as the speed at which the score of an attribute
          decreases over time.
    \item $T_{t}$ and $T_{t-1}$ are timestamps. $T_{t}$
          represents the current time and $T_{t-1}$ represents the last time
          this attribute received from a sightings. It is assumed that $T_{t} > T_{t-1}$.
\end{itemize}

\begin{figure}
\centering
    \includegraphics[scale=0.4]{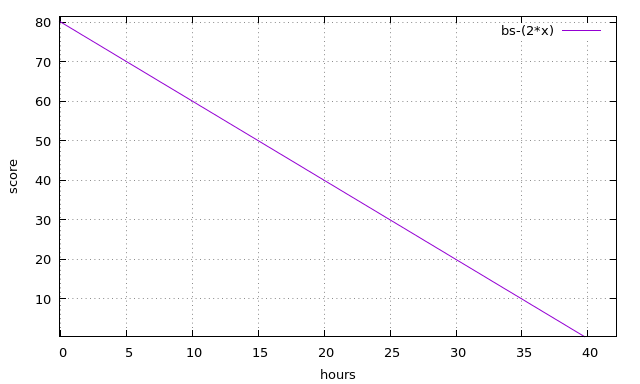}
    \caption{$\texttt{score}_a = \texttt{base\_score} -
              \delta_a (T_{t} - T_{t-1})$. The decay of the score is constant.}
    \label{fig:score1}
\end{figure}

Figure \ref{fig:score1} shows the decay of the score of an attribute with
a \texttt{base\_score} of $80$ and a decay rate $\delta_a$ of $2$.

An evaluation of the parameters shows that neither the end-time nor the
variable decay rate can be controlled.
Indeed, by fixing the decay rate, the end-time cannot be specified of the
score of an attribute. In the same mind, even if the decay rate is controlled
by the constant $\delta_a$, the decay is fixed over time.

To address the latter point, an exponential degression could be considered as
shown in equation \ref{score2}.

\begin{equation}\label{score2}
    \texttt{score}_a = base\_score_{a} \cdot e^{-\delta_a \cdot t}
\end{equation}

\begin{figure}
\centering
    \includegraphics[scale=0.4]{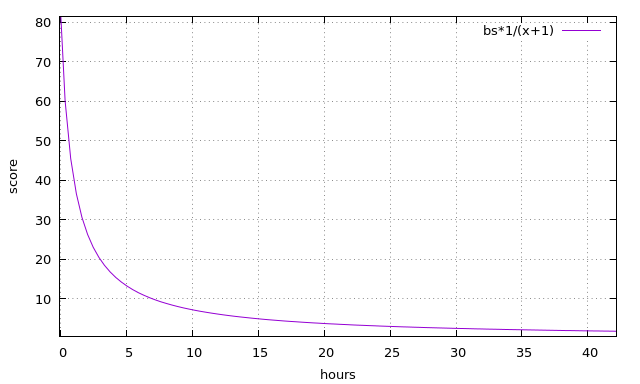}
    \caption{$\texttt{score}_a = \texttt{base\_score}
\cdot e^{-\delta_a \cdot t}$. The decay of the score follows an
exponential degression.}
    \label{fig:score2}
\end{figure}

In this case  a variable decay rate can be used. The slope in figure \ref{fig:score2}
is high at the beginning and lower as time passes.
However, the decay rate cannot be significantly influenced. This expression
cannot be used to have a slow decay at the beginning followed by a rapid
degression. A behavior that can, for example, be found in dynamic IP address
allocation by threat actors as previously described in this section.
Moreover, the time at which the overall score of the attribute should be 0
is entirely defined by the decay rate. So, manipulating the slope as well as
the end-time at the same time is still not possible.
Furthermore, it can be observed that the choice of the parameter $\delta_a$
will essentially range between 0 and 1 due to the tendency of the exponential
degression to rapidly tend to 0.

The final score is defined in equation \ref{score3}, capturing the conditions
stated previously.

\begin{equation}\label{score3}
    \texttt{score}_a = \texttt{base\_score} \cdot \left(1 - \left(\frac{t}{\tau_a}\right )^{\frac{1}{\delta_a}}\right )
\end{equation}
with
\begin{itemize}
    \item $\delta_a \in \left ] 0, +\infty \right.$, the decay speed.
    \item $\tau_a \in \left ] 0, +\infty \right.$, the end-time or time
needed such that \texttt{score}$_a = 0$. The end-time can be told by an
expiration sighting, where an organization knows when an indicator will be
expired. An example is the grace time: an Internet service provider gives a
grace time to customers to fix their machine until disconnecting them or
law enforcement agencies seizing the equipments.
It can also be derived from existing regular sightings, where organizations
provided data about sightings from the past.
    \item $t = T_t - T_{t-1}$, is an integer $>0$
\end{itemize}

\begin{figure}
\centering
    \includegraphics[scale=0.4]{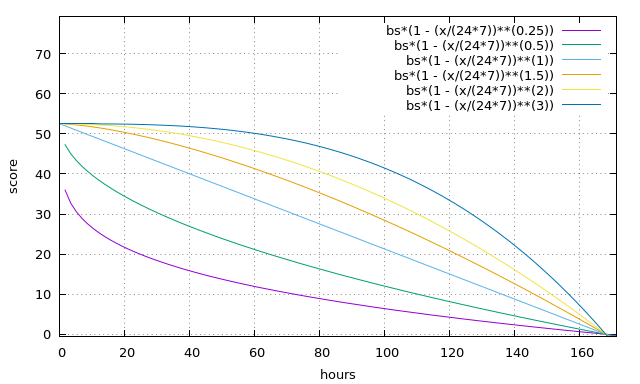}
    \caption{$\texttt{score}_a = \texttt{base\_score} \cdot \left(1 - \left(\frac{t}{\tau_a}\right )^{\frac{1}{\delta_a}}\right )$ for a fixed $\tau_a$ of 7 days}
    \label{fig:score3}
\end{figure}

This polynomial function has two advantages over the exponential one.
First, the end-time with $\tau_a$ can be easily controlled.
Second, the direction direction of the cavity of the slope can also be controlled.
A fast  degression at the beginning can be obtained
followed by a slow degression along with the complete opposite. An example
for a different decay rate $\delta_a$ can be seen in figure \ref{fig:score3}.
It can be seen that the greater $\delta_a$ is, the faster the overall score
decreases at the beginning. The more, the closer $\delta_a$ is to zero,
the slower the overall score will decrease at the start. 
The score is 0 for all decay rate for the specified $\tau_a$.

The parameters can be easily be fine tuned which is an additional advantage.
A value can be set for each type of attribute by performing a statistical
analysis on an existing dataset or users could set their own values via
a dedicated interface.

\begin{figure}
\centering
    \includegraphics[scale=0.4]{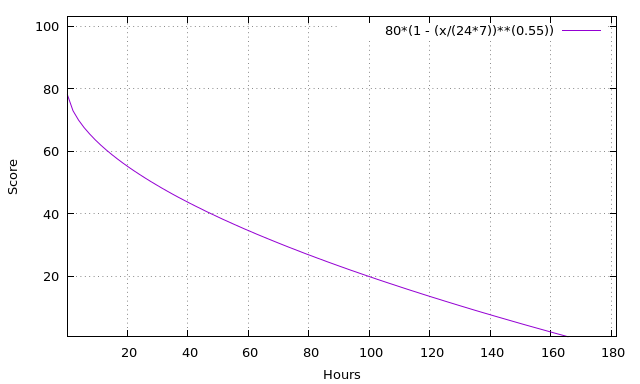}
    \caption{$\texttt{score}_a = 80 \cdot \left(1 - \left(\frac{t}{7\cdot 24}\right )^{\frac{1}{0.55}}\right )$ - It can be seen a rapid decrease of the score at the beginning. The score is halved after 48 hours.}
    \label{fig:ex1}
\end{figure}

Two examples are shown how the score in equation \ref{score3} can be used.
The first example is an attribute for
a compromised IP address being part of a botnet. The attribute of a shared
event in MISP belongs to the category {\tt Network activity} with its type
{\tt ip-dest}, meaning the destination IP address of a compromised webserver
hosting an exploitkit distributing malware. Some organizations spotted it
and started to share information about it. Abuse teams are informed to
cleanup the compromised systems. The IP address is encoded in publicly
available blacklists. The threat actors might notice the detection too
and start to move their exploitkit to another webserver. If we assume that
the Internet service provider gives a customer 1 week time to fix the
webserver. If it is not fixed within this time frame, the IP of the
webserver will be null-routed, meaning that it will not be accessible any more.
Hence, $\tau_a = 7\cdot 24$ hours. Under the hypothesis that the typical
blacklists take 48 hours to be applied in proxy servers or browsers,
the overall score should be halved after 2 days. Hence, $\delta_a = 0.55$.
Finally, if the base score of the attribute is calculated to be
\texttt{base\_score}$ = 80$ (based on the taxonomies and source confidence),
equation \ref{score3} becomes:

$$\texttt{score}_a = 80 \cdot \left(1 - \left(\frac{t}{7\cdot 24}\right )^{\frac{1}{0.55}}\right )$$
where $t$ is the time between now and the last \textit{sighting}, expressed
in hours. A plot of the decay is represented in figure \ref{fig:ex1}.

\begin{figure}
\centering
    \includegraphics[scale=0.4]{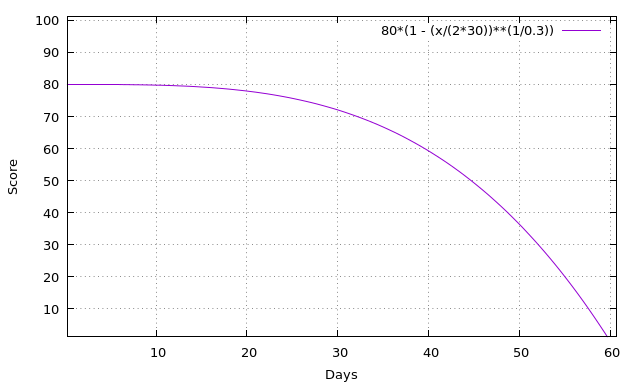}
    \caption{$\texttt{score}_a = 80 \cdot \left(1 - \left(\frac{t}{2\cdot 30}\right )^{\frac{1}{0.3}}\right )$ -  A really slow decrease at the beginning and a rush towards zero at the end can be observed. The overall score is halved in  48 days only.}
    \label{fig:ex2}
\end{figure}

The second example is the hash of a malware. If the hash is not a false positive
and a confirmed malware it says a malware and should not be decayed.
However, some host intrusion detection systems cannot handle million of hashes.
It could be considered
that the attribute will not have any value after 2 months, with a rather slow
decay if this is the expected time to destroy the attacking infrastructure.
$\tau_a = 2\cdot 30$ days and $\delta_a = 0.3$.
It is also supposed that the base score is the same as the previous example:
\texttt{base\_score}$= 80$. We have:
$$\texttt{score}_a = 80 \cdot \left(1 - \left(\frac{t}{2\cdot 30}\right )^{\frac{1}{0.3}}\right )$$
and the resulting plot can be seen in figure \ref{fig:ex2}.

\section{Future Work and Conclusions}
\label{conclusions}

This paper presents a work in progress for a scoring approach for
evaluating the decay methods of indicators of compromise that are shared
within threat intelligence platforms. The model is designed for the
peer to peer threat intelligence platform MISP taking into account existing
data annotations and data interaction methods.
Taxonomies attached to attributes are used in order to
get information about  reliability and confidence organizations have in
a given information source.
Each member organization can produce or consume
information, even consume information from multiple hops away. The
precondition in this case is that an information consumer trusts the
information producer.
Since the MISP platform grew organically from a handful interconnected
MISP instances, where members from the various communities knew and
trusted each other, to a large interconnected community having lots of
MISP instances in place. By this, nowadays, it is not unlikely that a
given information transits more MISP instances and only little
information known about the source and its producer.
This is a major drawback, because fake information can be shared for
harming an organization or to disrupt the sharing communities.
To counter this, another future work will be the investigation of
distributed information sharing with game theoretical approaches,
where detailed adversary models for peer to peer threat sharing will be studied.
Future research activities include the study of various models for
the $source\_confidence$. Here, evaluations will focus most probably on
the different machine learning techniques. For this, the model presented
in section \ref{sighting} in this paper has to be fully operational in order
to collect a data about sharing behaviours.

%
This work is co-financed by the European Union under the CEF grant 2016-LU-IA-0098.




\begin{thebibliography}{1}

\bibitem{Adams2011}
B.~Adams, D.Q.~Phung, and S.~Venkatesh.
\newblock Eventscapes: visualizing events over time with emotive facets.
\newblock In {\em Proceedings of the 19th International Conference on Multimedia 2011}, pages 1477--1480,Scottsdale, AZ, USA, 2011.ACM.

\bibitem{Barford}
P.~Barford, J.~Kline, D.~Plonka and A.~Ron.
\newblock A signal analysis of network traffic anomalies. ACM Sigcomm IMW, 2002.

\bibitem{Brown2015}
S.~Brown, J.~Gommers, and O.~Serrano.
\newblock From cyber security information sharing to threat management.
\newblock In {\em Proceedings of the 2Nd ACM Workshop on Information Sharing and Collaborative Security}, WISCS '15, pages 43--49, New York, NY, USA, 2015. ACM.

\bibitem{Pinto2015}
A.~Pinto, and A.~Sieira
\newblock Data‐Driven Threat Intelligence: Metrics on Indicator Dissemination and Sharing (\#ddti)
\newblock https://www.blackhat.com/docs/us-15/materials/us-15-Pinto-Data-Driven-Threat-Intelligence-Metrics-On-Indicator-Dissemination-And-Sharing.pdf, BlackHat 2015.

\bibitem{misppriv}
CIRCL.
\newblock Misppriv.
\newblock https://misppriv.circl.lu, 2016.

\bibitem{Dandurand13}
L.~Dandurand and O.~Serrano.
\newblock Towards improved cyber security information sharing.
\newblock In {\em Cyber Conflict (CyCon), 2013 5th International Conference on}, pages 1--16, 2013.

\bibitem{Dulaunoy}
A.~Dulaunoy and A.~Iklody.
\newblock MISP taxonomy format, https://tools.ietf.org/html/draft-dulaunoy-misp-taxonomy-format-00, 2016.

\bibitem{garcia}
S.~Garcia-Jimenez, E.~Magana, M.~Izal and D.~Moratoand.
\newblock IP addresses distribution in Internet and its application on reduction methods for IP alias resolution.
\newblock 34th Conference on Local Computer Networks. LCN 2009. IEEE.

\bibitem{Ghafouri}
A.~Ghafouri, W.~Abbas, A.~Lazka, Y.~Vorobeychik and X.~Koutsoukos.
\newblock Optimal Thresholds for Anomaly-Based Intrusion Detection in Dynamical Environments.
\newblock ArXiv e-prints 1606.06707, 2016.

\bibitem{Haass2015}
J.~C. Haass, G.-J. Ahn, and F.~Grimmelmann.
\newblock Actra: A case study for threat information sharing.
\newblock In {\em Proceedings of the 2Nd ACM Workshop on Information Sharing and Collaborative Security}, WISCS '15, pages 23--26, New York, NY, USA, 2015. ACM.

\bibitem{Enisa}
U.~Helmbrecht, S.~Purser, G.~Cooper, D.~Ikonomou, L.~Marinos, E.~Ouzounis, M.~Thorbrugge, A.~Mitrakas, and S.~Capogrossi.
\newblock Cybersecurity cooperation: Defending the digital frontline.
\newblock Technical report, ENISA, October 2013.

\bibitem{Prasad}
V.~Jyothsna and R.~Prasad.
\newblock A Review of Anomaly based Intrusion Detection Systems.
\newblock International Journal of Computer Applications, 2011.

\bibitem{blinc}
T.~Karagiannis, K.~Papagiannaki and M.~ Faloutsos
\newblock BLINC: Multilevel Traffic Classification in the Dark.
\newblock ACM SIGCOMM05, Philadelphia, Pennsylvania, USA, 2005.

\bibitem{Acing}
X.~Liao, K.~Yuan, X.~Wang, Z.~Liu and R.~ Beyah
\newblock Acing the IOC Game: Toward Automatic Discovery and Analysis of Open-Source Cyber Threat Intelligence.
\newblock 2016 ACM SIGSAC Conference on Computer and Communications Security, Vienna, Austria, 2016.

\bibitem{Maasberg2016}
M.~Maasberg, M.~Ko, and N.~L. Beebe.
\newblock Exploring a systematic approach to malware threat assessment.
\newblock In {\em 49th Hawaii International Conference on System Sciences (HICSS)}, pages 5517--5526, 2016.

\bibitem{misp-book}
MISP~Contributors.
\newblock User guide of misp malware information sharing platform, a threat
  sharing platform.
\newblock https://www.circl.lu/doc/misp/book.pdf, 2017.

\bibitem{Mitchell}
R.~Mitchell and I.R.~Chen.
\newblock A Survey of Intrusion Detection Techniques for Cyber-physical Systems.
\newblock ACM Comput. Surv., doi 10.1145/2542049, 2014.

\bibitem{Mun}
J.H.~Mun and H.~Lim.
\newblock New Approach for Efficient IP Address Lookup Using a Bloom Filter in Trie-Based Algorithms.
\newblock in IEEE Transactions on Computers, vol. 65, no. 5, pp. 1558-1565, May 1 2016.

\bibitem{Murdoch2015}
S.~Murdoch and N.~Leaver.
\newblock Anonymity vs. trust in cyber-security collaboration.
\newblock In {\em Proceedings of the 2Nd ACM Workshop on Information Sharing and Collaborative Security}, WISCS '15, pages 27--29, New York, NY, USA, 2015. ACM.

\bibitem{Nikulin}
V.~Nikulin.
\newblock Threshold-based clustering for intrusion detection systems.
\newblock Society of Photo-Optical Instrumentation Engineers (SPIE) Conference Series,  doi 10.1117/12.665326, 2006.

\bibitem{igen}
A.~Panwar.
\newblock iGen: Toward Automatic Generation and Analysis of IoCs using Convolutional Neural Networks.
\newblock https://repository.asu.edu/items/44216, 2017.

\bibitem{Dridex}
L.~Rudman and B.~Irwin.
\newblock Dridex: Analysis if the traffic and automatic generation of IOCs.
\newblock IEEE 2016 Information Security for South Africa (ISSA), 2016.


\bibitem{sommer}
R.~Sommer.
\newblock NetFlow: Information loss or win?
\newblock In {\em Proceedings of the 2nd ACM SIGCOMM Workshop on Internet measurement}, pp. 173-174, 2002.

\bibitem{straup2007machine}
A.~Straup~Cope.
\newblock Machine tags. flickr.
\newblock https://www.flickr.com/groups/api/discuss/72157594497877875/, 2007.

\bibitem{misp-article}
C.~Wagner, A.~Dulaunoy, G.~Wagener and A.~Iklody.
\newblock MISP: The Design and Implementation of a Collaborative Threat Intelligence Sharing Platform.
\newblock Proceedings of the 2016 ACM on Workshop on Information Sharing and Collaborative Security (WISCS'16), pages 49--56, 2016.
\newblock doi: 10.1145/2994539.2994542.

\bibitem{Waldvogel}
M.~Waldvogel, G.~Varghese J.~Turner and B.~Plattnery.
\newblock Scalable High Speed IP Routing Lookups.
\newblock In Proceedings of IEEE ACM SIGCOMM 97 Cannes, France, pp.25-36,
1997. 

\bibitem{Woods2015}
B.~Woods, S.~Perl, and B.~Lindauer.
\newblock Data mining for efficient collaborative information discovery.
\newblock In {\em Proceedings of the 2Nd ACM Workshop on Information Sharing and Collaborative Security}, WISCS '15, pages 3--12, New York, NY, USA,2015. ACM.


\end{thebibliography}
%

\end{document}